\documentclass[a4paper,11pt]{article}
\pdfoutput=1

\usepackage{jheppub}
\usepackage{amsmath,amssymb,amsfonts}
\usepackage{graphicx}
\usepackage{cleveref}
\usepackage{hyperref}
\usepackage{srcltx}
\usepackage{bm}
\usepackage{mathdots}
\usepackage{color}
\usepackage{dsfont}
\usepackage{slashed}
\usepackage{physics}
\usepackage{subfigure}
\usepackage{tikz}
\usepackage[normalem]{ulem}
\usepackage{orcidlink}
\usepackage{enumitem}
\usepackage{url}
\usepackage{comment}

\title{Multiple Axions Save High-Scale Inflation}

\author[a]{Dan Kondo,}\emailAdd{dan.kondo@ipmu.jp}
\author[a,b,c,1]{and Hitoshi Murayama\,\orcidlink{0000-0001-5769-9471}\note{Hamamatsu Professor}}\emailAdd{hitoshi@berkeley.edu}

\affiliation[a]{Kavli Institute for the Physics and Mathematics of the Universe (WPI), University of Tokyo Institutes for Advanced Study, University of Tokyo, Kashiwa 277-8583, Japan}
\affiliation[b]{The Leinweber Institute for Theoretical Physics, University of California, Berkeley, CA 94720, USA}
\affiliation[c]{Ernest Orlando Lawrence Berkeley National Laboratory, Berkeley, CA 94720, USA}

\abstract{
Many models of dark matter QCD axion require inflation at a scale $H_{\text{inf}} \lesssim 10^{6}$~GeV, which precludes a detectable tensor mode fluctuation. This constraint arises because the domain wall problem forces the Peccei--Quinn symmetry to be broken during inflation and the axions to be produced by the misalignment mechanism. We point out that theories with multiple axions can evade this constraint, thereby allowing high-scale inflation with detectable tensor mode. The only requirement is a condition on the anomaly coefficients that ensures a unique minimum of the axion potential without fine-tuning or the introduction of small parameters.} 

\begin{document}
\maketitle

\section{Introduction}

The Standard Model (SM) of particle physics is highly successful, but it leaves several puzzles unresolved. They include the identity of dark matter, the baryon asymmetry of the universe, and cosmic inflation. In particular, the axion~\cite{Weinberg:1977ma,Wilczek:1977pj} is one of the attractive candidates of dark matter that simultaneously addresses the strong CP problem~\cite{Peccei:1977hh}. A global Peccei--Quinn $U(1)_{PQ}$ symmetry is assumed to be spontaneously broken at a scale $f_a$ and produces the axion as its Nambu--Goldstone boson, while the QCD makes $U(1)_{PQ}$ anomalous and induces a small mass for the axion $m_a \simeq m_\pi^2 / f_a$.

On the other hand, inflation~\cite{Starobinsky:1980te,Guth:1980zm,Sato:1981ds,Linde:1981mu,Albrecht:1982wi} solves the flatness and horizon problems in cosmology, and creates the primordial density fluctuation by quantum fluctuation of the inflaton field. The current cosmological data strongly support this paradigm (see, {\it e.g.}\/, \cite{Planck:2018jri}). In particular, models with high inflationary scale $H_{\text{inf}}$ are attractive because they can leave observable imprints of the quantum fluctuation of the graviton, namely the ``quantum gravity" effects, via the B-mode polarization of the cosmic microwave background (CMB). These effects may be probed by forthcoming experiments such as CMB-S4 \cite{Abazajian:2019eic} or LiteBIRD \cite{delaHoz:2025uae}. 

In most discussions of axion dark matter, the misalignment mechanism \cite{Preskill:1982cy,Abbott:1982af,Dine:1982ah} is invoked. In this scenario, $U(1)_{PQ}$ is already broken during inflation, while the axion field is away from the true minimum during inflation held fixed by the Hubble friction. Only after inflation is over and the QCD-induced axion potential becomes important, the axion field starts to oscillate around the true minimum, making it a cold dark matter. 

Unfortunately, realizing both high-scale inflation and axion via the misalignment mechanism is difficult. Quantum fluctuations exist not only for the inflaton but also for the axion field, inducing unacceptably large isocurvature fluctuations unless $H_{\text{inf}} \lesssim 10^{-5} f_a$. Although models of low-scale inflation can be constructed, they sacrifice the ability to experimentally test inflation with CMB B-mode. If $f_a$ is much larger during inflation, it may suppress the isocurvature fluctuation sufficiently even when the inflationary scale is high \cite{Harigaya:2015hha}, though this entails significant challenges \cite{Graham:2025iwx}.

On the other hand, it is possible that $U(1)_{PQ}$ is spontaneously broken after inflation. In this case, axions are produced from axion strings and do not have isocurvature fluctuations, hence compatible with high-scale inflation. However, the symmetry breaking may lead to a network of domain walls \cite{Sikivie:1982qv}, a cosmological disaster. Such walls form when a discrete subgroup ${\mathbb Z}_{N_{DW}}$ of $U(1)_{PQ}$ remains exact, and different regions of the universe settle into different ground states via the Kibble--Zurek mechanism  \cite{Kibble:1976sj,Zurek:1985qw,Murayama:2009nj}. It is therefore essential to avoid degenerate ground states to allow for axion dark matter and high-scale inflation at the same time.

It turns out that many well-motivated models of axion have such an exact discrete symmetry. Most notably, Dine--Fischler--Srednicki--Zhitnitsky (DFSZ) axion models \cite{Zhitnitsky:1980tq,Dine:1981rt} have the standard model fermions charged under $U(1)_{PQ}$ and hence the number of vacua $N_{DW}$ is a multiple of three (number of generations). It is unfortunate given that DFSZ models are compatible with grand unification and hence are well motivated. 

There are various proposals to avoid the domain wall problem even when $U(1)_{PQ}$ is broken after inflation. So-called hadronic axion or Kim--Shifman--Vainshtein--Zakharov (KSVZ) models of axion \cite{Kim:1979if,Shifman:1979if} with only one color triplet quarks can avoid the problem. More recently, flavor-dependent assignment of $U(1)_{PQ}$ was proposed \cite{Ema:2016ops} which may also avoid domain walls \cite{Cox:2023squ} if charge assignments are chosen carefully. If the discrete symmetry is a part of the gauge group, the domain walls can be bound by strings, shrink and disappear \cite{Lazarides:1982tw}, which, however, often lead to other problems such as fractionally charged particles, Landau poles, etc \cite{Lu:2023ayc}. Explicit breaking of $U(1)_{PQ}$ can lift the degeneracy and make domain walls collapse \cite{Sikivie:1982qv}, but often requires a fine-tuning to maintain the quality of $U(1)_{PQ}$ to solve the strong CP problem \cite{Kawasaki:2014sqa}. 

In this paper, we consider the possibility that effective domain wall number can be one even for DFSZ axion. This is the case when there are multiple axions dubbed axiverse \cite{Svrcek:2006yi,Arvanitaki:2009fg}. Even when the anomaly coefficient for the QCD axion is larger than one such as for DFSZ axion models, mixing with another Axion-Like-Particle (ALP) makes the ground state unique and eliminates stable domain walls. Connection between multiple axions and domain walls was recently discussed with extensive numerical simulations in \cite{Benabou:2023npn,Lee:2024toz}. On the other hand, they did not discuss the possibility of testing high-scale inflation. In addition, we find their counting of ground states to be not useful for our purpose. We have identified the criteria for multi-axion potentials to have a unique minimum to avoid the domain wall problem.\footnote{We do not address the axion quality problem in this paper. See, {\it e.g.}\/, a recent attempt to solve the problem where $U(1)_{PQ}$ is an accidental symmetry of a composite dynamics \cite{Gherghetta:2025fip}. The proposal in this paper can in principle be combined with a composite dynamics.}

\section{DFSZ Axion and Domain Walls}

In this section, we briefly review the DFSZ axion and its domain wall problem. 

The original DFSZ axion models assigned the same $U(1)_{PQ}$ charges to all quarks and leptons to be consistent with the grand unified theories.\footnote{We do not necessarily require grand unification or supersymmetry, but we find it convenient and instructive to discuss explicit models in this language, and supersymmetry makes grand unification plausible because of the apparent unification of gauge coupling constants.} Clearly, there are many more possibilities to avoid domain walls beyond examples we present in this section.

The superpotential with two Higgs doublets is
\begin{align}
    W = Y_u Q U H_u 
    + Y_d Q d H_d + Y_l L E H_d + \lambda S H_u H_d,
\end{align}
where all matter chiral superfields $Q, U, D, L, N, E$ carry $Q_{PQ}=\frac{1}{2}$, both Higgs doublets $Q_{PQ} = -1$, and $S(+2)$.\footnote{Typically we need $\bar{S}(-2)$ to write a superpotential so that they acquire expectation value, {\it e.g.}\/, $W_S = Z (S \bar{S} - f_a^2)$ with $\langle S \rangle = \langle \bar{S} \rangle = f_a$.} The assumption is that $\mu = \lambda \langle S\rangle = \lambda f_a \simeq v$ where $v$ is the electroweak scale which requires $\lambda \ll 1$. This is a common issue with the DFSZ axion model which we do not discuss further. The axion field resides in these fields as $S = f_a e^{2ia/f_a}$, $H_u = v \sin\beta e^{-i a/f_a}$, $H_d = v \cos\beta e^{-i a/f_a}$. The anomaly coefficient is $N_{DW}=6$ and there is a domain wall problem.  This has been the source of the argument why ``GUT axion'' cannot be produced post inflation.

\section{Domain Wall Number With Multiple Axions}

In this section, we demonstrate that the number of degenerate ground states in a multi-axion potential is determined by the determinant of the matrix of anomaly coefficients. 

Domain walls arise in axion models when there is an exact discrete symmetry in the potential. Define $\theta = a/f_a$ which lives on $T^1$ where $\theta \simeq \theta + 2\pi$ are identified. When the axion potential takes the form\footnote{We are aware that the actual axion potential from the chiral Lagrangian is much more complicated. But for the purpose of identifying the minima, this simplified potential suffices. }
\begin{align}
    V = \Lambda^4 ( 1-\cos n \theta), \qquad
    n \in \mathbb{Z},
\end{align}
where $n$ is the anomaly coefficient for the gauge group. The potential has an exact $\mathbb{Z}_n$ discrete symmetry
\begin{align}
    \theta \rightarrow \theta + \frac{2\pi}{n} m, 
    \qquad m = 0, 1, \cdots, n-1.
\end{align}
As a result, the minima of the potential are at
\begin{align}
    \theta = \frac{2\pi}{n} m, \qquad
    m = 0, 1, \cdots, n-1.
\end{align}
Thus, there are $n$ independent minima of the potential for $\theta \in T^1$, when $n>1$. If this potential develops by a phase transition at some critical temperature $T_c$, the Kibble--Zurek mechanism \cite{Kibble:1976sj,Zurek:1985qw,Murayama:2009nj} leads to a network of domain walls because different parts of the universe choose different vacua that are separated by domain walls.

We can ask the same question when there are two axions and two potentials,
\begin{align}
    V &= V_1 + V_2, \\
    V_1 &= \Lambda_1^4 \left[1-\cos\left(n_{11} \theta_1 + n_{12} \theta_2\right)\right], \\
    V_2 &= \Lambda_2^4 \left[1-\cos\left(n_{21} \theta_1 + n_{22} \theta_2\right)\right].
\end{align}
The integers $n_{ij}$ are anomaly coefficients. The two angles live on a two-dimensional torus $(\theta_1, \theta_2) \in T^2$ due to their identifications $\theta_i \simeq \theta_i + 2\pi$ for both $i=1,2$. The question is then how to determine whether the potential possesses multiple minima on $T^2$. 

Obviously, the minima are given at\footnote{If one cares about the phase from quark mass one can regard them as a shift. One instead solves the equation $\cal{N}\vec{\theta}$$=2\pi\vec{\mathbb{Z}}-\vec{\delta}$, where $\vec{\delta}$ represents the shift. One can naturally obtain the minimum.}
\begin{align}
    n_{11} \theta_1 + n_{12} \theta_2 &= 2\pi m_1, \\
    n_{21} \theta_1 + n_{22} \theta_2 &= 2\pi m_2,
\end{align}
where $m_1, m_2 \in \mathbb{Z}$.  The above conditions can be cast to a matrix form,
\begin{align}
    {\cal N}
    \left( \begin{array}{c}
        \theta_1 \\
        \theta_2
        \end{array} \right)
    = \left( \begin{array}{cc}
        n_{11} & n_{12} \\
        n_{21} & n_{22} 
        \end{array} \right)
    \left( \begin{array}{c}
        \theta_1 \\
        \theta_2
        \end{array} \right)
    = 2\pi \left( \begin{array}{c}
        m_1 \\ m_2
        \end{array} \right).
        \label{eq:minima}
\end{align}
Here and below, we refer to the matrix ${\cal N}$ as the ``anomaly matrix.'' The basis vectors of the torus are mapped to
\begin{align}
    \vec{\theta}_1 = 2\pi \left( \begin{array}{c} 1 \\ 0 \end{array} \right)
    \rightarrow \vec{n}_1 = 2\pi \left( \begin{array}{c} n_{11} \\ n_{21} \end{array} \right), \qquad
    \vec{\theta}_2 = 2\pi \left( \begin{array}{c} 0 \\ 1 \end{array} \right)
    \rightarrow \vec{n}_2 = 2\pi \left( \begin{array}{c} n_{12} \\ n_{22} \end{array} \right).
\end{align}
The number of solutions is given by the area of the parallelogram on the $(m_1,m_2)$ plane spanned by these two vectors
\begin{align}
    |\vec{n}_1 \times \vec{n}_2| = |n_{11} n_{22} - n_{21} n_{12}|
    = |{\rm det} {\cal N}|.
\end{align}
This is the number of minima on the torus and hence the true ``domain wall number'' is $N_{DW} = |{\rm det} {\cal N}|$. See Section~\ref{sec:appendix} for the proof. 

\begin{figure}
    \centering
    \includegraphics[width=0.49\linewidth]{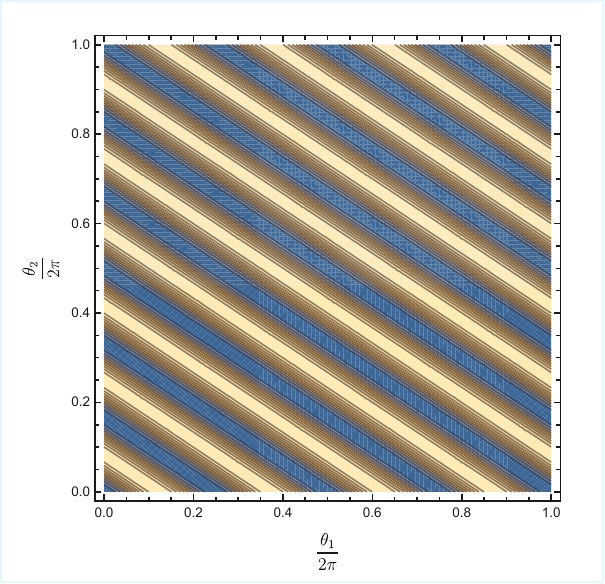}
    \includegraphics[width=0.49\linewidth]{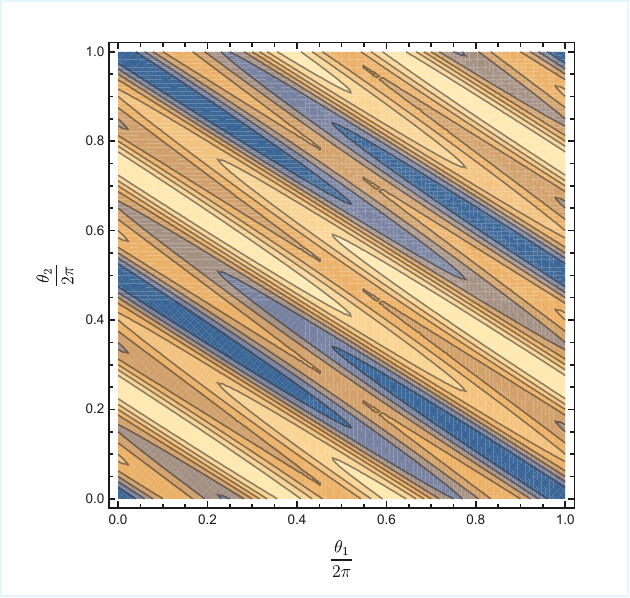}    \caption{The contour plots of the two-axion potential with ${\cal N}$ given in \eqref{eq:N0}. On the left plot, only $V_1$ is turned on. The minima are ${g_1 = \rm gcd}(4,6)=2$ disconnected lines. Even though the domain wall number ${\rm det}{\cal N}/g_1=1$ according to definitions in \cite{Benabou:2023npn,Lee:2024toz}, there are ${\rm det}{\cal N}=2$ degenerate minima after turning on $V_2$ at $(\theta_1, \theta_2) = (0,0)$ and $(0, \pi)$. Here, we took $\Lambda_1 = \Lambda_2 = 1$.}
    \label{fig:counterexample}
\end{figure}

In Refs. \cite{Benabou:2023npn,Lee:2024toz}, a smaller number is stated
\begin{align}
    N_{DW} = \frac{{\rm det}{\cal N}}{g_1}, \qquad
    g_1 = {\rm gcd}(n_{11}, n_{12})\ ,
\end{align}
where gcd refers to the greatest common divisor. This is somewhat misleading and requires a context to understand. In fact, even when this definition of $N_{DW}=1$, there are stable domain walls if $g_1>1$. 

They assumed a hierarchy $\Lambda_1 \gg \Lambda_2$, and first obtained minima for $V_1$. Then they counted the number of minima along the line $n_{11} \theta_1 + n_{12} \theta_2 = 0$, which is indeed ${\rm det}{\cal N}/g_1$. However, when $g_1 > 1$, there are disconnected lines that minimize $V_1$
\begin{align}
    n_{11} \theta_1 + n_{12} \theta_2 = 2\pi m_1, \qquad
    m_1 = (0, 1, \cdots, g_1 - 1).
    \label{eq:disconnected}
\end{align}
This can be seen as follows. B\'ezout's identity says there are sets of two integers $(\ell_1,\ell_2)$ that satisfy
\begin{align}
    n_{11} \ell_1 + n_{12} \ell_2 = g_1.
\end{align}
Therefore, the lines for
\begin{align}
    n_{11} \theta_1 + n_{12} \theta_2 = 0
\end{align}
and
\begin{align}
    n_{11} \theta_1 + n_{12} \theta_2 = 2\pi g_1
\end{align}
are related by the shift
\begin{align}
    \theta_1 \rightarrow \theta_1 + 2\pi \ell_1, \qquad
    \theta_2 \rightarrow \theta_2 + 2\pi \ell_2. 
\end{align}
The two lines are identified. However, since both $n_{11}$ and $n_{12}$ are multiples of $g_1$, there are no other sets of integers $(\ell_1, \ell_2)$ that allow shift in $m_1$ smaller than $g_1$. Hence $g_1$ lines in \eqref{eq:disconnected} are disconnected (left plot in Fig.~\ref{fig:counterexample}). Given that in each line there are ${\rm det}{\cal N}/g_1$ minima, and there are $g_1$ disconnected lines, the total number of minima is indeed ${\rm det}{\cal N}$ (right plot in Fig.~\ref{fig:counterexample}). For the illustration in the figure, we used
\begin{align}
    {\cal N}=\left( \begin{array}{cc} 4 & 6\\1 & 2 \end{array} \right).
    \label{eq:N0}
\end{align}

If $V_1$ has chosen one of the $g_1$ lines during the inflation, namely $\Lambda_1 \lesssim T_{RH}$, the remaining number of vacua is indeed $|{\rm det}{\cal N}/g_1|$, called pre-post-inflation scenario in \cite{Lee:2024toz}. However in this case the axion is produced via the misalignment and is again subject to the isocurvature perturbation. To accommodate high-scale inflation, one instead requires the post-post scenario, in which the $U(1)\times U(1)$ breaking occurs after inflation.

\section{Cosmological Evolution}

This analysis on the domain wall number in the previous section indicates that the cosmological evolution of topological defects is richer than what was discussed in~\cite{Lee:2024toz} when $g_1 > 1$.

When $U(1)_1 \times U(1)_2$ break sequentially, there are two independent networks of cosmic strings. At this point, any point on the torus is a minimum of the potential and different horizons choose different points. When $V_1$ turns on, domain walls form between the strings. If ${\rm min}(|n_{11}|,|n_{12}|) = 1$, a domain wall always ends with a string, forming the ``string bundle''~\cite{Eto:2023aqr}, since the domain wall tension eventually pulls the strings closer and the defect ends up being a closely held parallel bundle of strings that effectively behaves as a standard axion string.

If ${\rm min}(|n_{11}|,|n_{12}|) > 1$, every string is attached to multiple walls and is pulled in different directions. They can form a more complicated network that may not easily simplify. In other words, the network becomes frustrated, analogous to an anti-ferromagnet on a triangular lattice. Their simulation indeed showed that such a network can survive until $V_2$ turns on. When ${\rm det}{\cal N}/g_1 > 1$, the system devolves into stable domain walls which is a cosmological disaster. We agree with this picture. 

However, what is missing here is the possibility that separate infinite domain walls that do not end on pre-existing strings also form when $V_1$ turns on because of $g_1$ disconnected lines on the torus. These walls are stable even after $V_2$ turns on. This occurs even when ${\rm det}{\cal N}/g_1 = 1$. If ${\rm det}{\cal N}/g_1 > 1$, $V_2$ creates even more disconnected degenerate minima and hence domain walls (called ``induced domain walls'' in \cite{Lee:2024toz}).

When ${\rm det}{\cal N}=1$, there are no stable domain walls. If ${\rm min}(|n_{11}|,|n_{12}|) = 1$, the string bundles with $V_1$ behave as an axion string for the single axion case, which is filled by a domain wall once $V_2$ turns on, and collapse quickly. On the other hand if ${\rm min}(|n_{11}|,|n_{12}|) > 1$ all strings become attached to multiple walls and the system becomes frustrated. The string-wall network cannot simplify easily. But once $V_2$ turns on, there is a single minimum of the potential and the network does collapse.  

Here is an order of magnitude estimate of the condition for the collapse. Let us assume a hierarchy $\Lambda_1 \gg \Lambda_2$. To minimize $V_1$, we find the heavy and light (massless at this stage) axion  mass eigenstates, $\phi_H$ and $\phi_L$, respectively,
\begin{align}
    \phi_H &= \frac{1}{\sqrt{\left( \frac{n_{11}}{f_1} \right)^2 + \left( \frac{n_{12}}{f_2} \right)^2}} 
    \left(\frac{n_{11}}{f_1} a_1 + \frac{n_{12}}{f_2} a_2\right), \\
    \phi_L &= \frac{1}{\sqrt{\left( \frac{n_{11}}{f_1} \right)^2 + \left( \frac{n_{12}}{f_2} \right)^2}} 
    \left(\frac{n_{12}}{f_2} a_1 - \frac{n_{11}}{f_1} a_2\right) .
\end{align}
Then the potential is
\begin{align}
    V_1 = \Lambda_1^4 \left[ 1 - \cos \sqrt{\left( \frac{n_{11}}{f_1} \right)^2 + \left( \frac{n_{12}}{f_2} \right)^2}\ \phi_H \right],
\end{align}
and the effective decay constant is
\begin{align}
    f_H &= \frac{1}{\sqrt{\left( \frac{n_{11}}{f_1} \right)^2 + \left( \frac{n_{12}}{f_2} \right)^2}}\ .
\end{align}
To simplify the discussion, let us assume $f_1 \ll f_2$, so that $f_H \simeq f_1 / n_{11}$. Then the walls formed after $V_1$ has a tension $\sigma \approx \Lambda_1^2 f_H \simeq \Lambda_1^2 f_1$ (assuming $n_{11} = O(1)$ for the estimate). Therefore the mass of the network is $M \simeq \sigma H^{-2} \simeq \Lambda_1^2 f_1 H^{-2}$ within a horizon size $H^{-1}$. Once $V_2$ turns on, the wall receives a force due to the difference in the potential energies $F = \Delta V H^{-2} \simeq \Lambda_2^4 H^{-2}$. The acceleration is $a = F / M \simeq \Lambda_2^4 / (\Lambda_1^2 f_1)$. The walls collapse if their speed reaches the speed of light within the horizon time $a H^{-1} \gtrsim 1$, and hence
\begin{align}
    \Lambda_2^4 > \Lambda_1^2 f_1 H.
    \label{eq:collapse1}
\end{align}
We need the network to collapse well before the Big-Bang Nucleosynthesis, and hence $H \gtrsim T_{\rm coll}^2/M_{Pl} \simeq {\rm 10~MeV}^2/M_{Pl}$ ($T_{\rm coll}$ is the temperature at the time of the collapse). We know $\Lambda_2 = \Lambda_{QCD} \simeq 100$~MeV. We obtain a constraint
\begin{align}
    \Lambda_1^2 f_1 < \frac{M_{Pl}}{({\rm 10~MeV})^2} \Lambda_2^4
    \simeq 2 \times 10^{18}~{\rm GeV}^3.
    \label{eq:collapse2}
\end{align}

For instance, the following parameters saturate the constraint
\begin{align}
    \Lambda_1 \sim 10^5~{\rm GeV}, \quad
    f_1 \sim 10^{8}~{\rm GeV}, \quad f_2 \approx f_a \sim 10^{11}~{\rm GeV}.
\end{align}
Then the mass of the heavy ALP is
\begin{align}
    m_H \simeq \frac{\Lambda_1^2}{f_1} \simeq 10^2~{\rm GeV},
\end{align}
and it decays quickly into $gg$. There is no concern about its abundance or its contribution to $N_{\it eff}$. We will explore the parameter space fully in Section~\ref{sec:phenomenology}. 

If the model is supersymmetric, there is a potential issue with saxion dominating the energy density and producing a large entropy diluting baryon asymmetry and dark matter abundance. In our scenario, both axion sectors initially thermalize and their corresponding saxions are at the origin. Assuming second-order phase transitions, saxions adiabatically move to the final minima and we do not expect a problem. Obviously details depend on the dynamics of phase transitions and there are cases that may require further studies. We leave it for a later work. We repeat here that supersymmetry is optional in our discussion to enable grand unification and solve the hierarchy problem, not a neceessity.

\section{Generalized Supersymmetric DFSZ Axion}

\begin{table}[]
    \centering
    \begin{tabular}{|c|c|c|c|c|}
     \hline
         &$\mathrm{SU}(5)$& $\mathrm{U}(1)_{PQ}$ &$\mathrm{SU}(N_S)$ &$\mathrm{U}(1)_S$ \\\hline
        $10$ & 10& 0&1 &0\\ \hline
        $5^*$&5$^*$&1&1&0\\ \hline
        $N$ &1&-1&1&0\\ \hline
        $H_u$&5&0&1&0\\ \hline
        $H_d$&$\bar{5}$&-1&1&0\\ \hline
        $5'$&5&0&1&1\\ \hline
        $5^{*'}$&$\bar{5}$&0&1&1\\ \hline
        $\psi$&1&1/2&$N$&0\\ \hline
        $\psi'$&1&1/2&$\bar{N}$&0\\ \hline
        $\chi$&1&0&$N$&1/2\\ \hline
        $\chi'$&1&0&$\bar{N}$&1/2\\ \hline
    \end{tabular}
    \caption{Particle contents of our generalized DFSZ axion model with an additional $U(1)_S$ global symmetry and a strongly-coupled $SU(N_S)$ gauge group. This particle content leads to the anomaly matrix ${\cal N}$ given in \eqref{eq:N1}. We assume the usual triplet-doublet splitting so that we have only doublets below the GUT-scale in $H_{u,d}$.} 
    \label{tab:particlecontents}
\end{table}

In this section, we present a simple extension of the DFSZ axion model that avoids the domain wall problem through the introduction of an extra ALP. Clearly, there are many more possibilities beyond the example we present in this section.

We assign $U(1)_{PQ}$ differently from the original DFSZ model while maintaining the compatibility with $SU(5)$ grand unification. In addition, given that we have discovered tiny but finite neutrino masses, we most likely need right-handed neutrinos $N$ for the Type-I seesaw mechanism,
\begin{align}
    W = Y_u Q U H_u + Y_N L N H_u
    + Y_d Q d H_d + Y_l L E H_d + \lambda X_1 H_u H_d
    + \frac{1}{2} X_2 N N.
\end{align}
We assign charges shown in Table~\ref{tab:particlecontents}, where we used the $SU(5)$ notation $10 = (Q, U, E)$ and $5^* = (L, D)$. 
Here, $X_1$ and $X_2$ carry $Q_{PQ} = +1$ and $+2$, respectively, and corresponding fields $\bar{X}_{1,2}$ with opposite charges, and are assumed to have similar expectation values $\simeq f_2 \simeq f_a$. A generic superpotential such as
\begin{align}
    W_X =  M_1 X_1 \bar{X}_1 + M_2 X_2 \bar{X}_2 + X_2 \bar{X}_1^2 + \bar{X}_2 X_1^2
\end{align}
accomplishes this goal if $M_1 \simeq M_2$. 

At this stage, the anomaly coefficient for $U(1)_{PQ}$ is 3, and hence the model would lead to unacceptable domain walls. This feature is typical of DFSZ-type axion models. On the other hand, additional matter fields with standard-model (or $SU(5)$) quantum numbers are full $SU(5)$ multiplets and do not modify successful grand unification already known in supersymmetric standard model. 

Now we introduce a second axion (ALP) for a $U(1)_S$ symmetry, as well as a strongly-coupled $SU(N_S)$ gauge group. $\psi$ and $\chi$ ($\psi'$ and $\chi'$) are fundamentals (anti-fundamentals) of $SU(N_S)$, and they have global (anomalous) $U(1)_{PQ}$ and $U(1)_S$ charges. $U(1)_S$ symmetry is broken at a scale $f_1 \ll f_2$. For concreteness, we can introduce $S_1, S_2$ with $U(1)_S$ charges $+1/2$ and $1$, and correspondingly $\bar{S}_{1,2}$ similarly to the $X$ fields above. The $SU(N_S)$ gauge group becomes strong at $\Lambda_1 \gg \Lambda_2 = \Lambda_{QCD}$. 

\begin{figure}
    \centering
    \includegraphics[width=0.5\linewidth]{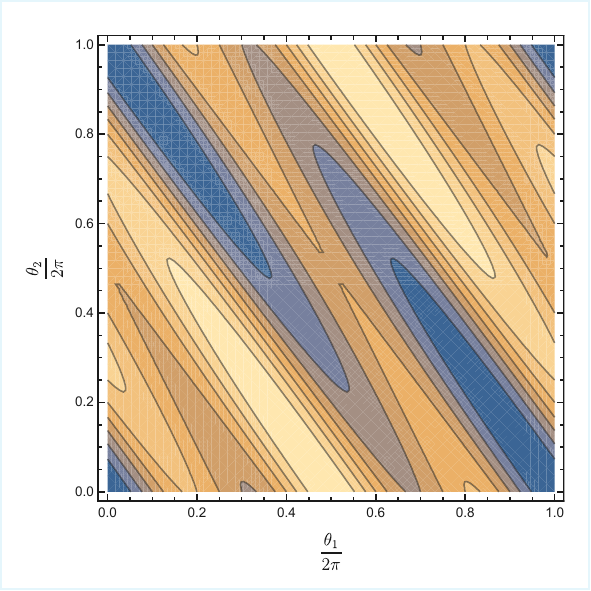}
    \caption{The plot of the potential which realizes $N_{DW}=1$ with ${\cal N}$ given in \eqref{eq:N1} taking $\Lambda_1 = \Lambda_2$ to make the minimum clearly visible. We can see that the minimum is unique at $(\theta_1, \theta_2) = (0,0)$.}
    \label{fig:plot1132V12}
\end{figure}

It is easy to see that the anomaly matrix for this particle content is
\begin{align}
    {\cal N} = \left( \begin{array}{cc} 1 & 1 \\ 3 & 2  \end{array} \right). \label{eq:N1}
\end{align}
Since ${\rm det} {\cal N} = -1$, there is a unique minimum for the potential. Since $n_{11}=n_{12}=1$, $V_1$ makes the strings into string bundles and cosmology reduces to that of a single QCD axion with $N_{DW}=1$ (Fig.~\ref{fig:plot1132V12}). The resulting axion abundance is expected to be the same as that of the standard axion string \cite{Buschmann:2021sdq},
\begin{align}
    \Omega_a \approx 0.12 h^{-2} \left( \frac{f_a}{4.3 \cdot 10^{10}~{\rm GeV}} \right)^{1.17} .
\end{align}
See the original reference for residual parameter dependences. 

\begin{figure}
    \centering
    \includegraphics[width=0.5\linewidth]{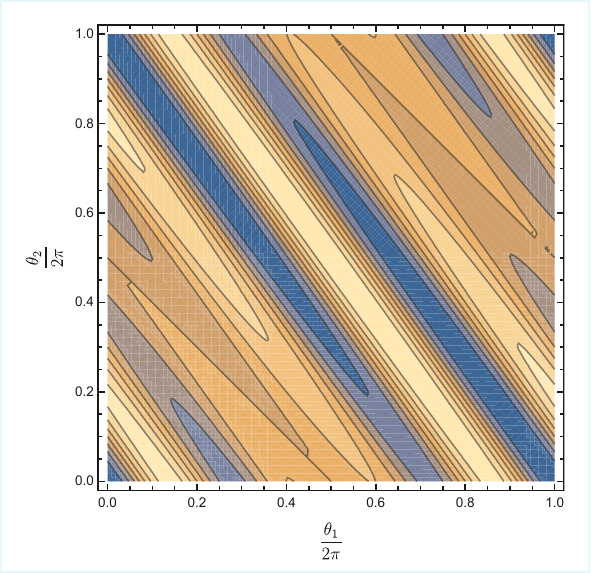}
    \caption{The plot of the potential which realizes $N_{DW}=1$ with ${\cal N}$ given in \eqref{eq:N2} taking $\Lambda_1 = \Lambda_2$ to make the minimum clearly visible. We can see that the minimum is unique at $(\theta_1, \theta_2) = (0,0)$.}
    \label{fig:plot4332V12}
\end{figure}

The model can be modified to allow for a string-wall network below $\Lambda_2$ that collapses below $\Lambda_{QCD}$ and may potentially lead to gravitational wave \cite{Lee:2024toz} that can explain the NANOGrav data \cite{NANOGrav:2023gor}. For instance, we can simply increase the number of $\psi+\psi'$ to 4 and $\chi+\chi'$ to 3, leading to the anomaly matrix
\begin{align}
    {\cal N} = \left( \begin{array}{cc} 4 & 3 \\ 3 & 2  \end{array} \right). \label{eq:N2}
\end{align}
Since ${\rm min}(n_{11},n_{12}) = 3$, the strings end up in a string-wall network that persists down to $\Lambda_{QCD}$. Yet ${\rm det}{\cal N} = -1$ (Fig.~\ref{fig:plot4332V12}) and hence the network collapses. This collapse at late times lead to the purported gravitational wave signature.

Note that there can be also a term $W=\lambda \bar{X}_1 5' 5^*$ that mixes the standard model fermions with the extra $5'+5^{\prime *}$. The mixing allows the particles at the Peccei--Quinn scale to decay into the standard-model particles.\footnote{If the mixing is sizable, it can allow the unsuccessful fermion mass relation to be relaxed and/or proton decay to be suppressed. We will not discuss these possibilities further in this paper.}

\section{Phenomenology} \label{sec:phenomenology}

In this section, we discuss the phenomenology of our multi-axion scenario in more detail. We heavily rely on numerical analyses in~\cite{Lee:2024toz}. In all cases, we assume both $U(1)$s are broken after inflation and lead to string networks called ``post-post inflation'' scenario. There are two different cases when $V_1$ at $\Lambda_1 \gg \Lambda_{2} = \Lambda_{QCD}$ turns on depending on whether (1) ${\rm min}(n_{11},n_{12}) = 1$ or (2) ${\rm min}(n_{11},n_{12}) > 1$. In case (1), strings form ``string bundle'' where walls end on a string, and strings are quickly are pulled together by the wall tension to effectively form a single-axion string. When $V_2$ turns on, the remaining string gets filled with a domain wall that quickly disappears. In case (2), strings are attached to multiple domain walls and the string-wall network becomes frustrated. When $V_2$ turns on, there is only one true minimum of the potential, the remaining domain walls move due to the potential difference between their two sides, collapse and disappear. We will discuss each case separately below.

\subsection{String Bundle ${\rm min}(n_{11},n_{12})=1$}

In this section, we discuss the phenomenology of the scenario.
We show the plot of the potential for this case in~\cref{fig:plot1132V12} taking the anomaly matrix \eqref{eq:N1}.

The phenomenological constraints on the parameter space are shown in~\cref{fig:DWpheno}.
\begin{figure}
    \centering
    \includegraphics[width=0.8\linewidth]{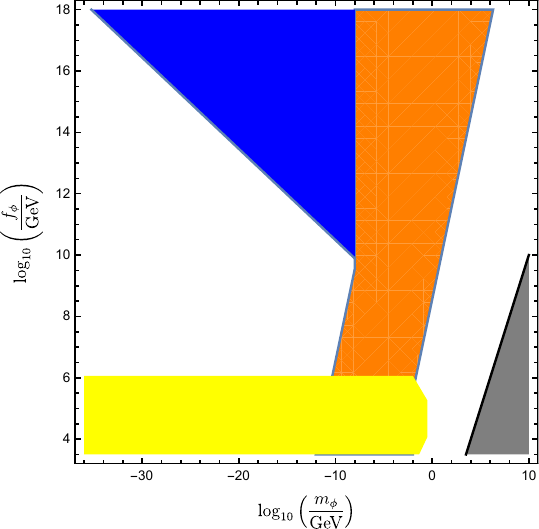}
    \caption{The phenomenology of the potential with a ``string bundle'' for $f_a=10^{11}$GeV as an example. The white region is the viable parameter space. The middle section in orange is excluded because ALPs decay late and destroys the success of the Big-Bang Nucleosynthesis \cite{Cadamuro:2011fd}. The lower left region in yellow is excluded because the excessively rapid cooling of SN1987A \cite{Cadamuro:2011fd}. The lower right region in dark gray is excluded because it theoretically does not make sense for an ALP if $m_\phi > f_\phi$. 
    }
    \label{fig:params}
\end{figure}

When the domain wall number is one ($N_{DW}=1$), the string-wall network collapses soon after formation due to the wall tension. The amount of the axion from the string and domain wall collapse are given by~\cite{Kawasaki:2014sqa,Buschmann:2021sdq} 
\begin{align}
    \Omega_{\text{string}}h^2&=7.3\times 10^{-3}\left(\frac{f_a}{10^{10}\text{GeV}}\right)^{1.18}\left(\frac{\Lambda}{400\text{MeV}}\right)\\
    \Omega_{\text{DW}}h^2&=3.7\times 10^{-3}\left(\frac{f_a}{10^{10}\text{GeV}}\right)^{1.18}\left(\frac{\Lambda}{400\text{MeV}}\right)
\end{align}

We show the parameter space of ALP $(m_\phi,f_\phi)$ in~\cref{fig:params}. We can see that there are safe regions free from domain wall problem. 

The constraint comes from the BBN, the decay of ALP can alter the baryon to photon ratio and change the D/H abundance~\cite{Cadamuro:2011fd}. The allowed parameter region is
\begin{align}
    m_\phi&>10\text{MeV}\ \text{and}\ \tau_{\phi\gamma\gamma}<10^{-2}\text{sec},\\
    m_\phi&<10\text{eV}\ \text{or}\ 
    \tau>10^{24}\text{sec}.
\end{align}
The excluded parameter region is shown in~\cref{fig:params} as the middle orange region.

The lifetime of ALP is approximately given by
\begin{align}
    \tau_{\phi\rightarrow gg}&=\left(\frac{\alpha_s}{32\pi^3}\frac{m_{\phi}^3}{f_\phi^2}\right)^{-1} 
    =4.9\times 10^{-2}~{\rm sec} \left(\frac{\alpha_s}{0.12}\right)^{-1}\left(\frac{m_{\phi}}{1\text{MeV}}\right)^{-3}\left(\frac{f_\phi}{10^5\text{GeV}}\right)^2, \\
        \tau_{\phi\rightarrow \gamma\gamma}&=\left(\frac{\alpha}{32\pi^3} \frac{m_{\phi}^3}{f_\phi^2}\right)^{-1}
        =0.82~{\rm sec}  \left(\frac{\alpha}{1/137}\right)^{-1}\left(\frac{m_{\phi}}{1\text{MeV}}\right)^{-3}\left(\frac{f_\phi}{10^5\text{GeV}}\right)^2 .
\end{align}
The requirement that the string-wall network collapses is given by Eq.~\eqref{eq:collapse2}, and is shown as the red line. 

Recall that we assume a post-post inflation scenario and hence the reheating temperature needs to be above $f_{1,2}$. The maximum reheating temperature is achieved by the instantaneous conversion of the inflation energy to radiation, where the former is constrained by the tensor mode fluctuation. As a result, there is an upper limit on the reheating temperature $T_{RH}<3\times10^{15}\text{GeV}$ and hence on the decay constants 
\begin{align}
    f_{1,2}< T_{RH} < 3 \times 10^{15}~{\rm GeV}.
    \label{eq:RH}
\end{align}
This is shown as the orange solid line in Fig.~\ref{fig:params}.

\subsection{String-Wall Network ${\rm min}(n_{11},n_{12})>1$}

When the generation of the dark sector quark is more than one, a non-zero domain wall number is expected. If we prepare for 4$\psi$'s and 3$\chi$'s, we have the anomaly matrix \eqref{eq:N2} with $n_{11}=4$, $n_{12}=3$. We show the potential in~\cref{fig:plot4332V12}. 

\begin{figure}
    \centering
    \includegraphics[width=0.8\linewidth]{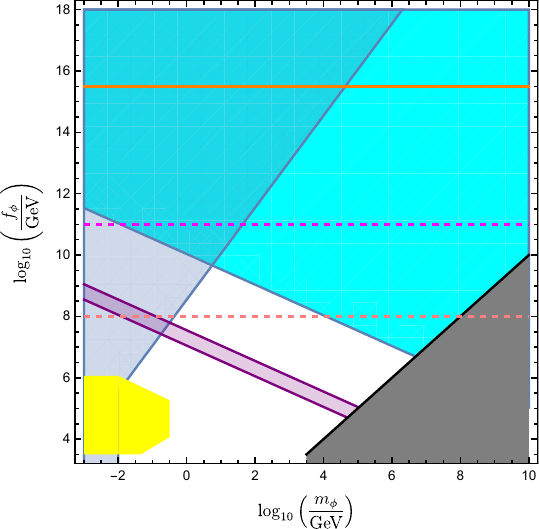}
    \caption{The phenomenology of the potential with a string-wall network for $f_a=10^{11}$GeV as an example. The white region is the viable parameter space. The upper left region in blue-gray is excluded because ALPs decay late and destroys the success of the Big-Bang Nucleosynthesis \cite{Cadamuro:2011fd}. The lower left region in yellow is excluded because the too-rapid cooling of SN1987A \cite{Cadamuro:2011fd}. The upper right region in cyan is excluded because the string-wall network persists down to the BBN era \eqref{eq:collapse2}. The lower right region in dark gray is excluded because it theoretically does not make sense for an ALP if $m_\phi > f_\phi$. Finally, the region below the orange dashed line is preferred to avoid axion overproduction from the collapse of string-wall network \eqref{eq:fratio}. We also show other implications for a reference. The orange line is the upper bound on decay constant from the tensor mode \eqref{eq:RH}. The magenta dashed line shows the $f_a$ and the pink dashed line represents $10^{-3}f_a$ to avoid the overproduction of axion DM from DW collapse. The magenta dashed line is $f_a = 10^{11}$~GeV for a comparison and we assumed $f_\phi < f_a$ and hence is below this line. The purple band roughly corresponds to the region where PTA signal can be explained ($10^{15}$GeV$^3\lesssim \sigma \lesssim 10^{16}$GeV$^3$)~\cite{Lee:2024xjb}.
    }
    \label{fig:DWpheno}
\end{figure}

The phenomenological constraints on the parameter space are shown in~\cref{fig:DWpheno}.

In this case, the domain walls collapse but walls are predominantly composed of ALPs. The mass of the network within the horizon size $H^{-1}$ is
\begin{align}
    M \approx \Lambda_1^2 f_1 H^{-2},
\end{align}
and the energy density is
\begin{align}
    \rho \approx \Lambda_1^2 f_1 H.
\end{align}
At the time of the collapse $H \simeq \Lambda_2^4 / \Lambda_1^2 f_1$ \cref{eq:collapse1}, 
\begin{align}
    \rho \approx \Lambda_2^4.
\end{align}
Since the thickness of the wall is of the order of $m_H^{-1} \simeq f_1 / \Lambda_1^2$, the ALPs are created only moderately relativistic $k \simeq m_H^{-1}$ \cite{Kawasaki:2014sqa} and quickly redshift to become cold. Namely the energy density of $\Lambda_2^4$ redshifts as matter and would quickly come to dominate the universe. This would be unacceptable. Therefore the ALPs need to decay quickly before the Big-Bang Nucleosynthesis. 

On the other hand, the collapse also has a small probability of producing axions suppressed by $\sim (f_1/f_2)^2$. The orange dashed line in Fig.~\ref{fig:DWpheno} is the upper limit on $f_1 \approx f_\phi$ to axion so that axions do not overclose the universe. When the domain wall collapses, the energy density of the domain wall is $\rho_{DW}\sim \Lambda_{QCD}^4=(0.3\text{GeV})^4$ and the entropy density is $s\sim 1.7\times 10^7 \text{MeV}^3 (g_*/10.8)(T_{QCD}/150\text{MeV})^3$. The ratio is
\begin{align}
    \frac{\rho_{DW}}{s}\simeq 5.0\times 10^2\text{MeV},
\end{align}
while the observed dark matter yield is $\rho_{\text{DM}}/s=0.44\text{eV}$. In order not to exceed the dark matter density, the ratio of the decay constant should satisfy
\begin{align}\label{eq:fratio}
    \frac{f_1}{f_2}<2.9\times 10^{-5}.
\end{align}

Interestingly, the collapse of the string-wall network can generate gravitational waves that may account for the NANOGrav signal \cite{NANOGrav:2023gor}. The preferred region \cite{Lee:2024xjb} for this is shown in the magenta band. 

\section{Conclusion}

In this paper, we proposed a new possibility: axion models with domain wall problems can be saved by having multiple axions, thereby making high-scale inflation compatible with such models. Once axion dark matter is discovered, $f_a$ and couplings to photons, electrons, nucleons will be measured \cite{Leedom:2025hpu}. In principle, these measurements would allow us to distinguish this scenario from the single axion case. At the same time, the observation of primordial CMB B-mode polarization would establish high-scale inflation which would render the misalignment mechanism unlikely. We anticipate significant progress from forthcoming experiments, including axion haloscope, axion helioscope, axion supernova-scope \cite{Ge:2020zww}, and CMB B-mode polarization. 

\section*{Note Added}

As we were finishing up this paper, another work appeared that also discusses solving domain wall problems with multiple axions \cite{Lee:2025zpn} by the same authors of Ref.~\cite{Lee:2024toz}. While their approach overlaps with ours, the motivations differ: they considered both pre- and post-inflationary ALPs, whereas our focus here is specifically on the post-inflationary scenario, which can render high-scale inflation viable. Another difference is our detailed discussion on $N_{DW}$, and explicit models that satisfy $N_{DW}=1$. 

\acknowledgments 
We thank Joshua Benabou, Ben Safdi, and Kai Murai for being helpful to understand their papers~\cite{Benabou:2023npn,Lee:2024xjb,Lee:2024toz}. D.\,K. gratefully thank the hospitality of DESY theory group during the preparation of this draft. D.\,K. was supported by JSPS KAKENHI Grant Number 24KJ0613. 
The work of H.\,M. is supported by the Director, Office of Science, Office of High Energy Physics of the U.S. Department of Energy under the Contract No. DE-AC02-05CH11231, by the NSF grant PHY-2210390, by the JSPS Grant-in-Aid for Scientific Research JP23K03382, MEXT Grant-in-Aid for Transformative Research Areas (A) JP20H05850, JP20A203, Hamamatsu Photonics, K.K, and Tokyo Dome Corportation. In addition, HM is supported by the World Premier International Research Center Initiative (WPI) MEXT, Japan. 

\appendix

\section{Explicit Proof for $N_{DW} = {\rm det}{\cal N} = 1$}
\label{sec:appendix}
When ${\rm det}{\cal N}=1$, there is only one minimum on the torus. We demonstrate this explicitly below. Now defining the greatest common divisors for both rows
\begin{align}
    g_1 &= {\rm gcd}(n_{11}, n_{12}), \\
    g_2 &= {\rm gcd}(n_{21}, n_{22}),
\end{align}
we can rewrite
\begin{align}
    \left( \begin{array}{cc}
        n_{11} & n_{12} \\
        n_{21} & n_{22} 
        \end{array} \right)
    = \left( \begin{array}{cc}
        g_1 p_1  & g_1 q_1 \\
        g_2 p_2  & g_2 q_2
        \end{array} \right),
\end{align}
where $p_1, q_1$ are relatively prime, and so are $p_2, q_2$. 

Then inverting \eqref{eq:minima}, we find
\begin{align}
    \left( \begin{array}{c}
        \theta_1 \\
        \theta_2
        \end{array} \right)
    &= 2\pi \frac{1}{g_1 g_2 (p_1 q_2 - p_2 q_1)}
        \left( \begin{array}{cc}
        g_2 q_2 & -g_1 q_1 \\
        -g_2 p_2 & g_1 p_1 
        \end{array} \right)
        \left( \begin{array}{c}
        m_1 \\ m_2
        \end{array} \right) \nonumber \\
    &= 2\pi \frac{1}{g_1 g_2 (p_1 q_2 - p_2 q_1)}
        \left( \begin{array}{cc}
        q_2 & -q_1 \\
        -p_2 & p_1 
        \end{array} \right)
        \left( \begin{array}{cc}
        g_2 & 0 \\
        0 & g_1 
        \end{array} \right)
        \left( \begin{array}{c}
        m_1 \\ m_2
        \end{array} \right) \nonumber \\
    &= 2\pi \frac{1}{g_1 g_2 (p_1 q_2 - p_2 q_1)}
        \left( \begin{array}{cc}
        q_2 & -q_1 \\
        -p_2 & p_1 
        \end{array} \right)
        \left( \begin{array}{c}
        g_2 m_1 \\ g_1 m_2
        \end{array} \right) .
\end{align}
The question is whether these minima are located elsewhere from $(\theta_1, \theta_2) = (0,0)$ on $T^2$. The question is whether the right-hand side yields integer multiples of $2\pi$ or fractional values. 

Consider $m_1=1$, $m_2 = 0$. Then
\begin{align}
    \left( \begin{array}{c}
        \theta_1 \\
        \theta_2
        \end{array} \right)
    &= 2\pi \frac{1}{g_1 g_2 (p_1 q_2 - p_2 q_1)}
        \left( \begin{array}{cc}
        q_2 & -q_1 \\
        -p_2 & p_1 
        \end{array} \right)
        \left( \begin{array}{c}
        g_2 \\ 0
        \end{array} \right) \nonumber \\
    &= 2\pi \frac{1}{g_1 (p_1 q_2 - p_2 q_1)}
        \left( \begin{array}{c}
        q_2 \\
        -p_2 
        \end{array} \right) .
\end{align}
If $g_1 (p_1 q_2 - p_2 q_1)=1$, the solution lies on the lattice point and is therefore equivalent to the origin. When $g_1 (p_1 q_2 - p_2 q_1)=k \neq 1$, at least either $\theta_1$ or $\theta_2$ is not on the lattice point because $p_2, q_2$ are relatively prime and cannot be divided by another integer $k$ to yield integers. Thus, additional minima are avoided if and only if $g_1 (p_1 q_2 - p_2 q_1)=1$. Exactly the same argument for $m_1=0$, $m_2=1$ requires $g_2 (p_1 q_2 - p_2 q_1)=1$. Both conditions can be satisfied only when
\begin{align}
    g_1 = g_2 = p_1 q_2 - p_2 q_1 =1.
    \label{eq:conditions}
\end{align}
Namely, $n_{11}, n_{12}$ must be relatively prime, $n_{21}, n_{22}$ also must be relatively prime, and $n_{11} n_{22} - n_{12} n_{21} = 1$. This is the requirement not to have any stable domain walls.

In fact, the discussion generalizes to larger number of axions in the following way. For a potential with an equal number $N$ of axions as the strong gauge groups,
\begin{align}
    V = \sum_{i=1}^N \Lambda_i^4 \left[ 1 - \cos \left( \sum_{j=1}^N n_{ij} \theta_j \right) \right],
\end{align}
the minima are found when
\begin{align}
    {\cal N} \vec{\theta} =
    \left( \begin{array}{ccc}
        n_{11} & \cdots & n_{1N} \\
        \vdots & \ddots & \vdots \\
        n_{N1} & \cdots & n_{NN}
        \end{array} \right)
    \left( \begin{array}{c}
        \theta_1 \\ \vdots \\ \theta_N
        \end{array} \right)
    = 2\pi \left( \begin{array}{c}
        m_1 \\ \vdots \\ m_N
        \end{array} \right)
    = 2\pi \vec{m},
\end{align}
where $m_j \in {\mathbb Z}$. The $N$-dimensional torus $T^N$ is mapped to each of the column vectors in ${\cal N}$ in the same of ${m_i}$, whose volume ${\rm det}{\cal N}$ is nothing but the number of minima. Conversely for each choice of $\vec{m}$, the minimum is determined to be at
\begin{align}
    \vec{\theta} = {\cal N}^{-1} \vec{m}.
\end{align}
When ${\rm det}{\cal N}=1$, ${\cal N}^{-1} = ({\rm cofactor}\ {\cal N})^T$, whose entries are all integers by definition. Therefore, for every $\vec{m}$, the minima are at $\theta_i \in 2\pi {\mathbb Z}$. Namely, ${\cal N} \in SL(N,{\mathbb Z})$ is a necessary and sufficient condition to have a unique minimum of the potential. 

\bibliographystyle{JHEP}
\bibliography{ref}
\end{document}